\documentstyle[sprocl]{article}

\input{psfig}

\def\Journal#1#2#3#4{{#1} {\bf #2}, #3 (#4)}

\def\PLB{{\em Phys. Lett.}  B}

\def\PRD{{\em Phys. Rev.} D}

\def\be{\begin{equation}}
\def\ee{\end{equation}}
\def\bea{\begin{eqnarray}}
\def\eea{\end{eqnarray}}

\begin{document}

\vspace*{-3.0 cm}
\begin{flushright}
CERN-TH/2000-169 
\end{flushright}
\vspace*{0.5 cm}

\title{LEPTON-FLAVOUR-VIOLATION IN SUSY MODELS\\ WITH AND
WITHOUT R-PARITY
\footnote{Talk given at the 8th International Workshop on Deep Inelastic
Scattering and QCD (DIS 2000), Liverpool, England, 25-30 April 2000}
}
\author{KAZUHIRO TOBE}

\address{CERN, Theory Division, CH-1211, Geneva 23, Switzerland
\\E-mail: kazuhiro.tobe@cern.ch
} 


\maketitle\abstracts{ 
We discuss Lepton-Flavour-Violating phenomena such as
$\mu \rightarrow e \gamma$, $\mu \rightarrow eee$, and
$\mu \rightarrow e$ conversion in nuclei in SUSY models with
and without R-parity. We stress that experimental searches for 
all the LFV processes are important to distinguish between the 
different models.}
Recently, the atmospheric neutrino experiment SuperKamiokande has
announced evidence for non-zero neutrino mass, which indicates
the existence of physics beyond the Standard Model (SM).
Generally in models that accommodate neutrino oscillations,
lepton-flavour-violating phenomena (LFV) can occur not
only in the neutrino, but also in the charged-lepton sector.
In this paper, we will discuss LFV in processes
such as $\mu \rightarrow e \gamma$, $\mu \rightarrow eee$, and
$\mu \rightarrow e$ conversion in nuclei in two models, which
are extensions of the minimal SUSY SM (MSSM):
one is the MSSM with right-handed neutrinos 
(without R-parity violation),\cite{LFV_nR} 
the other is the MSSM with R-parity violation.\cite{RPV} Especially the latter 
model can also be tested by the HERA experiments;\cite{HERA} 
however, the low energy muon-decay
experiments provide very stringent bounds on some of the couplings,
which are not constrained by HERA. 
We will also discuss the different features of LFV in these two models.

\vspace*{-0.1 cm}
\section{LFV in the MSSM with right-handed neutrinos}
\begin{figure}[t]
\psfig{figure=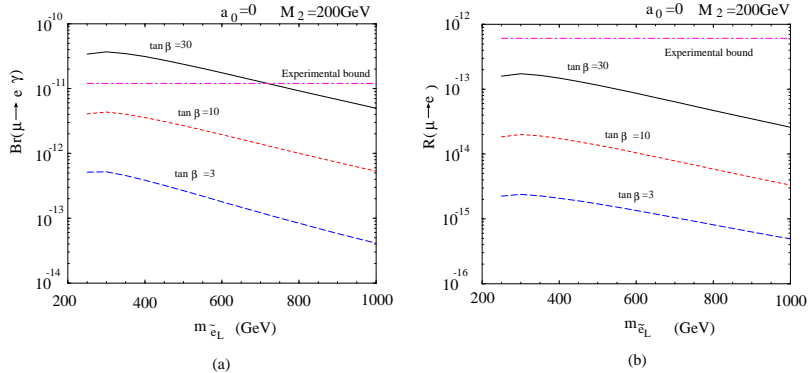,angle=-90,height=5cm}
\caption{Event rates for $(a)$ $\mu \rightarrow e \gamma$
$(b)$ $\mu \rightarrow e$ conversion in Ti as a function of
the left-handed selectron mass. We fixed the right-handed neutrino mass
to $M_R=10^{13}$ GeV.}
\label{LFV_muon}
\end{figure}

In this section, we consider the MSSM with right-handed neutrinos.
The superpotential in the lepton sector is given by
\begin{eqnarray}
W&=&f_e^{i} H_1 E^c_i L_i + f_\nu^{i} V_{ij} H_2 N^c_i L_j 
+\frac{M_R}{2} N^c_i N^c_i,
\end{eqnarray}
where for simplicity we assumed that the right-handed neutrino mass matrix
is proportional to the unit matrix. In this framework, the small neutrino 
masses arise through the seesaw mechanism; if the right-handed neutrino mass
scale $M_R$ is much larger than the electroweak scale, we can obtain
very tiny neutrino masses:
$m_{\nu i} = (m_{\nu i}^D)^2/M_R, ~~\nu_{\rm mass}^i = 
V_{ij} \nu_{\rm flavor}^j$,
where $m_\nu^D$ is a Dirac neutrino mass $(m_{\nu i}^D=
\frac{f_\nu^i v \sin\beta}{2})$, and the mass eigenstates ($\nu_{\rm mass}$)
of neutrinos are related to the flavour eigenstates ($\nu_{\rm flavor}$)
via the mixing matrix $V$. An important point is that,because of the LFV
mixing $V$, LFV is induced in the slepton mass terms
even if we assume universal scalar mass $(m_0)$ at the gravitational 
scale $M=2\times 10^{18} {\rm GeV}$. We can calculate the LFV in 
slepton masses by solving the renormalization group equations 
numerically. The approximate solution is given by
$(\Delta m^2_{\tilde{L}})_{ij} \simeq-\frac{|f_\nu^k|^2 V_{ki} V_{kj}}
{16 \pi^2} (6+2 a_0^2)m_0^2 \log (M/M_R)$.
Note that large neutrino Yukawa couplings and large mixing $V$
induce large flavour-violating masses.
The LFV masses $(\Delta m^2_{\tilde{L}})_{ij}$ generate LFV
processes,  e.g. $(\Delta m^2_{\tilde{L}})_{23}$
induces $\tau \rightarrow \mu \gamma$, and $(\Delta m^2_{\tilde{L}})_{12}$
generates $\mu \rightarrow e$ flavour violation.
The atmospheric neutrino experiments indicate large mixing $V_{32}$. This
large mixing can induce large branching ratio for
$\tau \rightarrow \mu \gamma$~\cite{LFV_nR,Hisano_Nomura}. 
Moreover, some of the solar neutrino solutions
suggest large $V_{21}$, which can induce large $\mu \rightarrow e$ flavour 
violation.\cite{Hisano_Nomura}
In Fig.~\ref{LFV_muon}, we present the branching ratios of 
$\mu \rightarrow e \gamma$ and
$\mu \rightarrow e$ conversion in Ti assuming that 
\begin{eqnarray}
V &=&\left(
\begin{array}{ccc}
\frac{1}{\sqrt{2}} & \frac{1}{2} & \frac{1}{2}\\
\frac{-1}{\sqrt{2}} & \frac{1}{2} & \frac{1}{2}\\
0 & \frac{-1}{\sqrt{2}} & \frac{1}{\sqrt{2}}
\end{array}
\right),~~
m_{\nu i} =(0,~0.006,~0.055)_i~{\rm eV},
\end{eqnarray}
which provides a solution to the atmospheric neutrino problem, as well as a 
large-angle MSW solution to the solar neutrino 
problem.\footnote{In the case with other solar neutrino solutions, 
see Ref.\cite{Hisano_Nomura}}
We also checked that 
$\rm{Br}(\mu \rightarrow eee)/\rm{Br}(\mu \rightarrow e \gamma)
\simeq 6\times 10^{-3}$.
Therefore the present and future experiments with 
a sensitivity of $10^{-14}$
for the $\mu \rightarrow e \gamma$ rate~\cite{PSI} and $10^{-16}$ for
the $\mu \rightarrow e$ conversion rate in Al~\cite{MECO}\footnote{
We numerically checked that the $R(\mu \rightarrow e~ {\rm in~ Al}) 
\simeq 0.6 R(\mu \rightarrow e~ {\rm in~ Ti})$.}
can probe the LFV in a large region of the parameter space of this model.

\vspace*{-0.1 cm}
\section{LFV in SUSY models with R-parity violation}
Subsequently, we consider the MSSM with R-parity violation in which the small
R-parity violation can potentially explain small neutrino masses.
The superpotential is
\begin{eqnarray}
W=f_e^i H_1 E^c_i L_i + \frac{f_{ijk}}{2} L_i L_j E^c_k +f'_{ijk}L_i Q_j D^c_k.
\end{eqnarray}
Here we assume that the baryon-number-violating terms $U^c D^c D^c$ 
are forbidden by baryon parity in order
to avoid rapid proton decay.\footnote{We neglect R-parity violation
in soft SUSY-breaking terms. This could be important for neutrino masses.
Here we do not consider any particular models for neutrino masses.}
The Yukawa couplings $f$ and $f'$ violate lepton flavour number as well as 
lepton number. The LFV experiments put the most stringent
bounds on certain combinations of the Yukawa couplings 
$f$ and $f'$.\cite{Choudhury,Huitu}
We list the bounds in Table~\ref{bound}. To understand an interesting 
feature of the LFV in this model, let us consider some simple examples. 
If only $f_{131}$ and $f_{231}$ are non-zero, the ratios of the event rates
between the LFV processes are given by
\begin{eqnarray}
\frac{{\rm Br}(\mu \rightarrow e \gamma)}{{\rm Br}(\mu \rightarrow eee)}
=1\times 10^{-4},~~\frac{R(\mu \rightarrow e~{\rm in~Ti~(Al)})}
{{\rm Br}(\mu \rightarrow eee)} = 2\times 10^{-3}~(1\times 10^{-3}),
\label{model1}
\end{eqnarray}
where we took  all $m_{\tilde{l}}$ to be $100$ GeV. The rate of
$\mu \rightarrow eee$ is much larger than those of the other 
LFV processes. This is because $\mu \rightarrow eee$ is generated at 
tree level, while the rest of the LFV processes are induced at the one-loop 
level. 
If only $f_{132}$ and $f_{232}$ are non-zero, the event rates are given by
\begin{eqnarray}
\frac{{\rm Br}(\mu \rightarrow e \gamma)}{{\rm Br}(\mu \rightarrow eee)}
=1.2,~~\frac{R(\mu \rightarrow e~{\rm in~Ti~(Al)})}
{{\rm Br}(\mu \rightarrow eee)} = 18~(11),
\label{model2}
\end{eqnarray}
where we took $m_{\tilde{l}}=100$ GeV. Even though all 
processes are induced at the one-loop level in this case, the branching
ratio of $\mu \rightarrow eee$ is still comparable to that of 
$\mu \rightarrow e \gamma$, and the rate of $\mu \rightarrow e$ conversion
is even much larger because of a log-enhancement in the off-shell photon
penguin contributions.\cite{Huitu} Therefore $\mu \rightarrow eee$ and
$\mu \rightarrow e$ conversion are important in this model.
The interesting point is that the relations in Eqs.(\ref{model1}),
(\ref{model2})
are quite different from those of the MSSM with right-handed neutrinos
discussed in the previous section. Therefore, in order to distinguish
between the different models, the study of all the LFV processes 
can be important. Note that at present
the future proposal for $\mu \rightarrow eee$ is missing.
Furthermore, we should stress here that not only low-energy muon 
probes but also other experiments\cite{HERA}
should be encouraged in order to investigate the R-parity-violating 
SUSY since this theory is well-motivated 
from the recent neutrino data.

\begin{table}[t]
\begin{center}
\begin{tabular}{|c|c|c|c|} \hline
 & $\mu \rightarrow e\gamma$ & $\mu \rightarrow 3e$ 
& $\mu \rightarrow e~{\rm in~nuclei}$
\\ \hline
$|f_{131}f_{231}|$ & 
$2\times 10^{-4}~(7\times 10^{-6})$ &$7\times 10^{-7}$
&$1\times 10^{-5}~(2\times 10^{-7})$
\\
$|f_{132}f_{232}|$ & 
$2\times 10^{-4}~(7\times 10^{-6})$& $7\times 10^{-5}$
 &$1\times 10^{-5}~(2\times 10^{-7})$ 
\\
$|f_{133}f_{233}|$ & 
$2\times 10^{-4}~(7\times 10^{-6})$& $1\times 10^{-4}$
 & $2\times 10^{-5}~(4\times 10^{-7})$
\\
$|f_{121}f_{122}|$ & 
$8\times 10^{-5}~(2\times 10^{-6})$ &$7\times 10^{-7}$
&$ 6\times 10^{-6}~(1\times 10^{-7})$
\\
$|f_{131}f_{132}|$ & 
$8\times 10^{-5}~(2\times 10^{-6})$ &$7\times 10^{-7}$
&$7 \times 10^{-6}~(1\times 10^{-7})$
\\
$|f_{231}f_{232}|$ & 
$8\times 10^{-4}~(2\times 10^{-6})$& $4\times 10^{-5}$ &
$8\times 10^{-6}~(1\times 10^{-7})$
\\ \hline
$|f'_{111}f'_{211}|$ & 
$7\times 10^{-4}~(2\times 10^{-5})$ & $1\times 10^{-4}$ &
$5\times 10^{-6}~(2\times 10^{-7})$
\\
$|f'_{112}f'_{212}|$ & 
$7\times 10^{-4}~(2\times 10^{-5})$ & $1\times 10^{-4}$ &
$4\times 10^{-7}~(7\times 10^{-9})$
\\
$|f'_{113}f'_{213}|$ & 
$7\times 10^{-4}~(2\times 10^{-5})$ & $2\times 10^{-4}$ &
$4\times 10^{-7}~(7\times 10^{-9})$
\\
$|f'_{121}f'_{221}|$ & 
$7\times 10^{-4}~(2\times 10^{-5})$ & $2\times 10^{-4}$ &
$4\times 10^{-7}~(6\times 10^{-9})$
\\
$|f'_{122}f'_{222}|$ & 
$7\times 10^{-4}~(2\times 10^{-5})$ & $2\times 10^{-4}$ &
$4\times 10^{-5}~(7\times 10^{-7})$
\\
$|f'_{123}f'_{223}|$ & 
$7\times 10^{-4}~(2\times 10^{-5})$ & $3\times 10^{-4}$ &
$5\times 10^{-5}~(9\times 10^{-7})$
\\
$|f'_{131}f'_{231}|$ & 
$2\times 10^{-3}~(6\times 10^{-5})$ & $4\times 10^{-4}$ &
$4\times 10^{-7}~(6\times 10^{-9})$
\\
$|f'_{132}f'_{232}|$ & 
$2\times 10^{-3}~(6\times 10^{-5})$ & $5\times 10^{-4}$ &
$9\times 10^{-5}~(2\times 10^{-6})$
\\
$|f'_{133}f'_{233}|$ & 
$2\times 10^{-3}~(6\times 10^{-5})$ & $9\times 10^{-4}$ &
$2\times 10^{-4}~(3\times 10^{-6})$
\\ \hline
\end{tabular}
\end{center}
\caption{Present (Future) constraints on R-parity-violating 
couplings from LFV processes. The present limits (future
expectations) on the event rates are given by
${\rm Br}(\mu \rightarrow e\gamma)<1.2\times10^{-11}~(10^{-14})$,
${\rm Br}(\mu \rightarrow 3e)<1.0\times 10^{-12}$, and
${\rm R}(\mu \rightarrow e~{\rm in~Ti~(Al)}) < 6.1\times 10^{-13}~(10^{-16})$.
We took $m_{\tilde{\nu}}=m_{\tilde{l}_R}=100$ GeV and $m_{\tilde{q}}=300$
GeV.}
\label{bound}
\end{table}

\vspace*{-0.3 cm}
\section*{Acknowledgements}
The author would like to thank A. de Gouv\^ea, G.F. Giudice, and S. Lola 
for useful discussions.
%
%
%

\end{document}